\documentclass[aps,prl,reprint,superscriptaddress,amsmath,amssymb]{revtex4-1}

\usepackage{graphicx}
\usepackage{hyperref}
\usepackage{physics}
\usepackage{siunitx}
\usepackage{color}
\usepackage{upgreek}
  
\graphicspath{../Figures}

\DeclareMathOperator{\di}{d\!}

\newcommand{\bit}{\begin{itemize}}
\newcommand{\eit}{\end{itemize}}

\newcommand{\bea}{\begin{eqnarray}}
\newcommand{\eea}{\end{eqnarray}}

\newcommand{\be}{\begin{equation}}
\newcommand{\ee}{\end{equation}}

\newcommand{\B}{\mathbf}
\newcommand{\figref}[1]{Fig.~\ref{#1}}

\renewcommand{\eqref}[1]{Eq.~(\ref{#1})}

\usepackage{soul}

\begin{document}

\title{Rotons in Optical Excitation Spectra of Monolayer Semiconductors}

\author{Ovidiu Cotlet}
\email{ocotlet@phys.ethz.ch}
\affiliation{Institute of Quantum Electronics, ETH Z{\"u}rich, CH-8093, Z\"urich, Switzerland }

\author{Dominik S.~Wild}
\email{wild@g.harvard.edu}
\affiliation{Department of Physics, Harvard University, Cambridge, Massachusetts 02138, USA}

\author{Mikhail D.~Lukin}
\affiliation{Department of Physics, Harvard University, Cambridge, Massachusetts 02138, USA}

\author{Atac Imamoglu}
\affiliation{Institute of Quantum Electronics, ETH Z{\"u}rich, CH-8093, Z\"urich, Switzerland }

\begin{abstract}
 Optically generated excitons dictate the absorption and emission spectrum of doped semiconductor transition metal dichalcogenide monolayers. We show that upon increasing the electron density, the elementary optical excitations develop a roton-like dispersion, evidenced by a shift of the lowest energy state to a finite momentum on the order of the Fermi momentum. This effect emerges due to Pauli exclusion between excitons and the electron Fermi sea, but the robustness of the roton minimum in these systems is a direct consequence of the long-range nature of the Coulomb interaction and the nonlocal dielectric screening characteristic of monolayers. Finally, we show that the emergence of rotons could be related to hitherto unexplained aspects of photoluminescence spectra in doped transition metal dichalcogenide monolayers.
\end{abstract}



\maketitle

  Rotons are quasiparticles whose dispersion exhibits a minimum at finite momentum. The concept was first proposed by Landau as an explanation for the properties of superfluid $^4$He~\cite{landau1941theory,henshaw1961modes}. More recently, rotons have been predicted and observed in ultracold Bose gases with long-range interactions~\cite{mottl2012roton,chomaz2018observation}, engineered spin--orbit coupling~\cite{khamehchi2014measurement,ji2015softening}, and shaken optical lattices~\cite{ha2015roton-maxon}. However, the study of rotons and other many-body effects in ultracold gases is impeded by the weak interactions between neutral atoms. It is therefore desirable to explore platforms with strong interparticle interactions that simultaneously retain a greater degree of tunability than conventional condensed matters systems such as $^4$He. Over the past few years, monolayers of semiconducting transition metal dichalcogenides (TMDs) have emerged as promising candidates in this respect~\cite{mak2016photonics}. TMDs support tightly bound neutral excitons with binding energies of several hundred meV, which facilitates optical investigation of many-body effects. The charge carrier density of TMDs can be readily tuned with electrostatic gates. It is further possible to modify the properties of TMDs by stacking them with other van der Waals materials into heterostructures~\cite{geim2013van} or by changing the dielectric environment~\cite{raja2017coulomb}. 

  In this Letter, we investigate the many-body states formed by excitons in doped TMD monolayers. For concreteness, we focus on MoSe$_2$, although we expect our results to qualitatively apply to other TMDs. We first consider a single exciton in a Fermi sea of electrons, where all electrons, including the one that makes up the exciton, are spin and valley polarized. This regime can be accessed experimentally with moderate magnetic fields~\cite{back2017giant,wang2018strongly}. The exciton dispersion is significantly altered by the Fermi sea since the constituent electron is restricted to states above the Fermi energy, $E_F$. We will show that as the Fermi energy is increased from zero, the binding energy of the exciton decreases while its mass increases. At a critical Fermi energy, the exciton mass diverges and the energy minimum is shifted to a finite momentum $p_\text{rot}$, which we refer to as a roton minimum.

\begin{figure}[t]
  \includegraphics[width=\columnwidth]{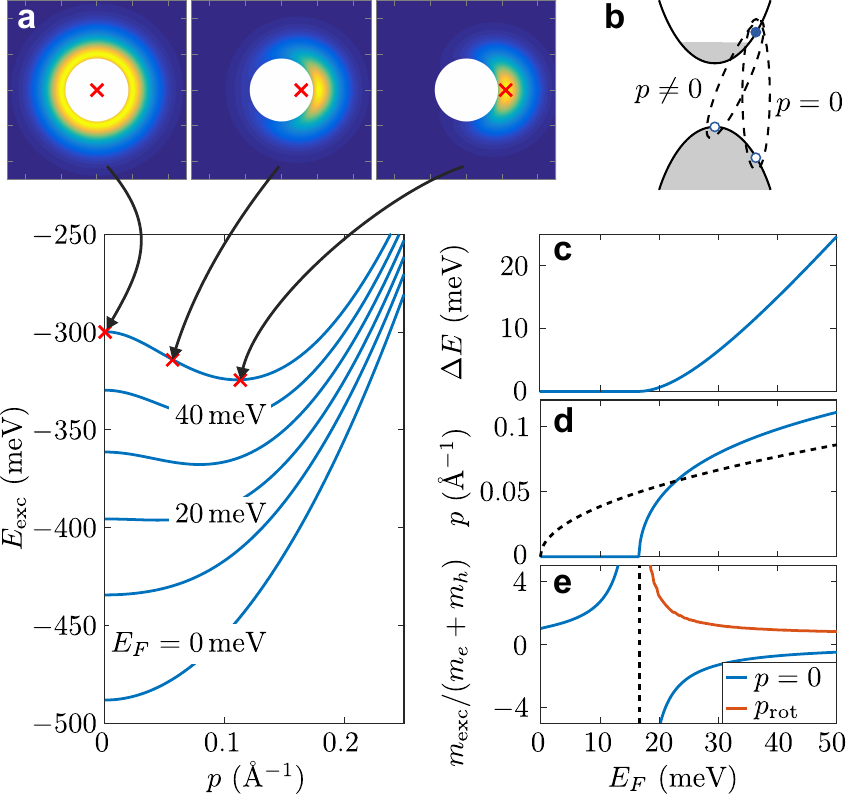}
  \caption{(a) Lower panel: Exciton dispersion at different Fermi energies, $E_F$. Upper panels: Electron wavefunction $|\varphi_{\B p}(\B k)|^2$ as a function of $(k_x, k_y)$ at three different center of mass momenta $\B p$ indicated by the red crosses. The blocked out circle is Fermi sea. The axis limits are $\pm \SI{0.25}{\angstrom^{-1}}$. (b) Schematic drawing illustrating the origin of the roton-like dispersion. As a function of Fermi energy: (c) Energy difference $\Delta E = E_\text{exc}(0) - \mathrm{min} \, E_\text{exc}(\B p)$ between the zero momentum exciton and the lowest energy state. (d) Momentum of the lowest energy exciton state (solid) and Fermi momentum (dashed). (e) Exciton mass at zero momentum (blue) and at the roton minimum (red). }
  \label{fig:fig1}
\end{figure}

  In contrast to the systems studied previously, this roton minimum is a consequence of Pauli blocking rather than interparticle interactions. As illustrated in Fig.~\ref{fig:fig1}(b), the exciton reduces the kinetic energy of the hole by adopting a nonzero center-of-mass momentum. A closely related effect has been predicted for the molecular state of an impurity in an ultracold Fermi gas, where the roton minimum is associated with the formation of FFLO states~\cite{massignan2014polarons}. These states have however never been observed as they are expected to be unstable over a wide range of parameters~\cite{parish2013highly}. At the same time, the roton minimum for excitons in TMDs is expected to be much more robust owing to the long-range nature of the Coulomb interaction compared to short-range interatomic interactions in ultracold gases.
  
The treatment of the fully polarized  regime serves as a starting point for the experimentally more relevant case where both valleys are populated. The exciton, taken to occupy the $K$ valley, now additionally interacts with the Fermi sea electrons in the opposite $K'$ valley. As a result, the exciton is dressed by electron-hole excitations in the $K'$ conduction band, which leads to the formation of new quasiparticles known as attractive and repulsive polarons~\cite{sidler2016fermi,efimkin2017many-body}. We will demonstrate that the attractive polaron inherits the roton-like dispersion from the exciton. This offers a possible explanation for the difference between absorption and photolumiscence (PL) spectra of MoSe$_2$ at finite doping~\cite{sidler2016fermi,wang2017probing,back2017giant}. 

  We now proceed to a quantitative discussion, starting with the fully polarized case. We are interested in the bound state between an electron in the conduction band and a hole in the valence band. The Hamiltonian of the electron-hole system $H_\text{exc} = H_K+H_I$ can be decomposed into a kinetic energy part $H_K$ and an interaction part $H_I$. The kinetic energy is simply $H_K=\sum_{\B k} \varepsilon_e(\B k)  e^\dagger_{\B k} e_{\B k}+\sum_{\B k} \varepsilon_h(\B k) h^\dagger_{\B k} h_{\B k}$, where $e^\dagger$ ($h^\dagger$) creates an electron in the conduction band (hole in the valence band) with the respective dispersion $\varepsilon_{e,h}(\B k) = |\B k|^2/(2 m_{e,h})$. The interaction Hamiltonian is given by the (attractive) Coulomb interaction between electrons and holes, $H_I=-\sum_{\B k \B{k'} \B q} V(\B q) e^\dagger_{\B k + \B q} h^\dagger_{\B{k'}-\B q}  h_{\B{k'}} e_{\B k}$, where we neglected electron-electron as well as electron-hole exchange terms~\cite{yu2014dirac,wu2015exciton,qiu2015nonanalyticity,glazov2014exciton,yu2014valley,glazov2017intrinsic}. In the above, $V(\B q)= 1/ [2\epsilon_0 \epsilon(\B q) | \B q| ]$ denotes the Coulomb interaction between electrons in a monolayer TMD, which is modified by nonlocal dielectric screening specific to monolayers, where  $\epsilon(\B q)=1+r_0 | \B q| $ and $r_0\approx \SI{5}{nm}$ in MoSe$_2$~\cite{berkelbach2013theory}. 

  We calculate the exciton energy variationally. Defining the exciton creation operator $x_{\B p}^\dagger = \sum_{\B k} \varphi_{\B p}(\B k) e^\dagger_{\B k} h^\dagger_{\B p - \B k}$ for a given total momentum $\B p$, we minimize $W_\text{exc}(\B p) = \bra{\text{FS}} x_{\B p} [H_\text{exc} - E_\text{exc}(\B p) ] x_{\B p}^\dagger \ket{\text{FS}}$, where $\ket{\text{FS}}$ denotes the electron Fermi sea. The electron making up the exciton is thus confined to states above the Fermi sea ($|\B k| > k_F$). The exciton energy $E_\text{exc}(\B p)$ (measured with respect to the semiconductor band gap at $E_F=0$) acts as a Lagrange multiplier to ensure the exciton state is suitably normalized. We solve the eigenvalue problem corresponding to $\delta W_{\text{exc}}(\B p) / \delta \varphi_{\B p}(\B k) = 0$ numerically using angular momentum eigenstates on a nonuniform radial grid (see \cite{SupMat}).

The dispersion thus obtained is shown in \figref{fig:fig1}(a) for Fermi energies between $E_F = \SI{0}{meV}$ and $E_F = \SI{50}{meV}$. At large doping, the dispersion minimum shifts to finite momentum. The schematic illustration in \figref{fig:fig1}(b) provides a qualitative explanation for the emergence of such a roton state. With the electron constrained to states above the Fermi sea, an electron-hole pair with zero total momentum carries a larger kinetic energy than an electron-hole pair in which the hole occupies a state closer to the valence band maximum. At small Fermi energies, the Coulomb attraction more than compensates for the cost in kinetic energy, allowing the lowest energy exciton to remain at zero momentum. The energy cost increases with increasing Fermi energy such that the excitonic ground state eventually shifts to finite momentum. While the roton minimum originates from Pauli blocking, the exact form of the electron-hole interaction plays a crucial role in determining the Fermi energy at which the roton minimum appears. In our case, using the dielectrically screened Coulomb potential, the roton minimum develops at $E_F \approx \SI{20}{meV} \approx E_0/25$, where $E_0$ denotes the exciton binding energy at zero Fermi energy. By contrast, a much greater Fermi energy of $E_F = E_0/2$ would be required if the electron and hole interacted via contact interaction (assuming equal carrier masses)~\cite{parish2013highly}. The Pauli blocking effect is more pronounced for long-range interactions because scattering with large momenta to accessible states across the Fermi sea is suppressed. It is especially efficient in monolayer materials due to the short-range dielectric screening which further suppresses momenta on the order of $1/r_0$.

The properties of the dispersion are explored more quantitatively in \figref{fig:fig1}(c)--(e). \figref{fig:fig1}(c) shows the energy separation between the exciton at zero momentum and the lowest energy state, indicating that the ground state is no longer at zero momentum when the Fermi energy exceeds $E_F \approx \SI{20}{meV}$. In \figref{fig:fig1}(d), we plot the momentum at which the dispersion minimum occurs. As expected from the qualitative argument given in the context of \figref{fig:fig1}(b), the momentum of the roton is on the order of the Fermi momentum (dashed line). Finally, the effective mass both at zero momentum and at the roton minimum, $p_\text{rot}$, is plotted \figref{fig:fig1}(e). The effective mass at the roton minimum corresponds to the curvature of the dispersion along the radial direction, while the mass in the tangential direction diverges as a consequence of rotational symmetry.

The above treatment ignores screening by the Fermi sea. A common approach to include screening for the exciton problem is based on the GW approximation and the corresponding screened ladder approximation ~\cite{gao2016dynamical,van2017marrying}. However, since the plasma frequency ($\omega_\text{pl}(\B q) = \sqrt{n_e |\B{q}|^2 V(|\B{q}|) / m_e} \approx \SI{90}{meV}$ at $|\B q| = k_F$ for a Fermi energy of $E_F = \SI{20}{meV}$) is much smaller than the exciton binding energy, the electron gas is unable to respond on the time scales characteristic for the electron-hole interaction. More specifically, the time scale for the bound electron--hole pair to scatter off each other via Coulomb interaction is set by $1/E_\text{exc} \ll 1 / \omega_\text{pl}$, implying that many scattering events may occur during the lifetime of a virtual plasmon. This is in stark contrast to the physical picture of the noncrossing, screened ladder approximation, in which virtual excitations of the Fermi sea are created and annihilated sequentially before the bound electron-hole pair scatters again. Therefore, the aforementioned treatment of screening is likely be valid only in the limit $1/E_\text{exc} \gg 1 / \omega_\text{pl}$, while in the opposite limit, the electrons might prefer to screen the exciton as a whole \cite{sidler2016fermi}. The above discussion highlights that an accurate treatment of screening is challenging and offers an explanation for the observed stability of excitons at high carrier densities~\cite{yao2017optically}, which is unexpected from the screened ladder calculations~\cite{van2017marrying}.

In our approach, screening can be taken into account by extending our ansatz to allow for electron--hole pair excitations in the Fermi sea. Such an ansatz not only enables a treatment of screening beyond the screened ladder approximation but it has moreover been shown to increase the robustness of the roton-like dispersion in ultracold Fermi gases~\cite{parish2011polaron-molecule,parish2013highly}. Translated to our system, the improvement of the variational ansatz is a consequence of the fact that in the original electronic wavefunction, the roton state carries a large electron momentum of order $p_\text{rot}$ despite its zero group velocity [see \figref{fig:fig1}(a)]. A wavepacket formed by states near the roton minimum would therefore disperse very quickly, which implies that our ansatz is far from an energy eigenstate. This argument was used by Feynman to motivate the introduction of backflow corrections in superfluid helium~\cite{feynman1956energy}, which significantly lower the variational energy of the roton. A quantitative treatment of these effects is however beyond the scope of this Letter. 

We next discuss the situation in which the electrons are unpolarized. Since there exists a trion bound state between the excitons in the $K$ valley and electrons in the $K'$, dressing of the exciton by the electrons in the opposite valley must be considered. We treat this problem within the rigid exciton approximation~\cite{sidler2016fermi,efimkin2017many-body}, in which the Hamiltonian is given by
  \bea\label{Heff}
  H_\text{pol} =\sum_{\B k} E_\text{exc}(\B k) x^\dagger_{\B k} x_{\B k} &+& \sum_{\B k} \varepsilon_e(\B k) e^{\prime\dagger}_{\B k} e_{\B k}^\prime \\
  &+& \sum_{\B k, \B k', \B q} U_{\B q} x^\dagger_{\B k+ \B q} e^{\prime \dagger}_{\B{k'}- \B q} e_{\B{k'}}^\prime x_{\B k} , \nonumber
  \eea
  where $E_\text{exc}(\B k)$ is the exciton dispersion computed above, $x_{\B k}^\dagger$ creates an exciton at momentum $\B{k}$ in the $K$ valley, and $e_\B{k}^{\prime \dagger}$ creates an electron at $\B{K'} + \B{k}$. We further assume a contact interaction potential between electrons and excitons, $U_{\B q} = U$, where the strength $U$ of the interaction can be related to the experimentally accessible trion binding energy $E_T$, i.e.~the binding energy of one exciton and one electron. It is possible to show from \eqref{Heff} that $ 1/U = - \sum_{|\B k|<\Lambda} [E_T+E_\text{exc}(\B k)+\varepsilon_e(\B k)]^{-1}$, at $E_F = 0$, where $\Lambda$ is a high-momentum cutoff that regularizes the interaction~\cite{parish2013highly}. For the remaining numerical calculations, we use $E_T = \SI{30}{meV}$~\cite{ross2013electrical} and employ a finite cutoff, $\Lambda = \SI{1}{nm^{-1}}$. The choice of cutoff is physically motivated by the exciton Bohr radius $a \approx \SI{1}{nm}$, which introduces a length scale at which the Hamiltonian in \eqref{Heff} becomes invalid as the exciton may no longer be treated as rigid, point-like object.

  \begin{figure}
    \includegraphics[width=\columnwidth]{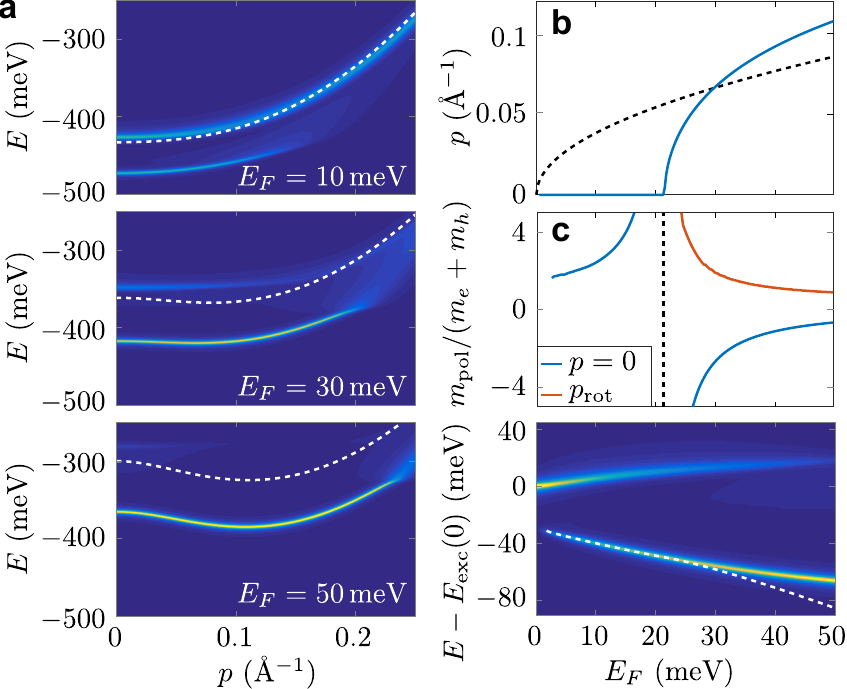}
    \caption{(a) Spectral function $S(E, \B p)$ for three different Fermi energies, $E_F$. The dashed line indicates the exciton dispersion $E_\text{exc}(\B p)$. As a function of Fermi energy: (b) Momentum of the lowest energy attractive polaron state (solid) and Fermi momentum (dashed). (c) Mass of the attractive polaron at zero momentum (blue) and at the roton minimum (red). (d) Optical conductivity $\sigma(E)$  according to \eqref{eq:conductivity}. The lowest energy of the attractive polaron is shown by the dashed line. A linewidth of $\gamma = \SI{3}{meV}$ was used.}
    \label{fig:fig2}
  \end{figure}

  The interaction with electrons gives rise to an exciton self-energy $\Sigma(E, \B p)$ and the dressed exciton (or Fermi polaron) energy $E_\text{pol}(\B p)$ can be obtained by solving the equation $E_\text{pol}(\B p) = E_\text{exc}(\B p) + \mathrm{Re} \, \Sigma(E_\text{pol}(\B p), \B p)$. We calculate the self-energy using the non-self-consistent ladder approximation of the $T$-matrix~\cite{efimkin2017many-body,schmidt2012fermi}. This approach is equivalent to a variational treatment known as the Chevy ansatz ~\cite{chevy2006universal}, where a polaron with momentum $\B p$ is associated with a creation operator of the form $b^\dagger_{\B p} = \alpha_{\B p} x^\dagger_{\B p} + \sum_{\B k, \B q} \beta_{\B p, \B k, \B q} x^\dagger_{\B p+ \B q- \B k} e^{\prime \dagger}_{\B k} e_{\B q}^\prime$. The polaron dispersion can then alternatively be computed by minimizing the quantity $W_\text{pol}(\B p) = \bra{\text{FS}} b_{\B p} [H_\text{pol} - E_\text{pol}(\B p)] b^\dagger_{\B p} \ket{\text{FS}}$. The result is shown in Fig.~\ref{fig:fig2}(a) at three different Fermi energies, where we plot the spectral function $S(E, \B p)= - 2 \, \mathrm{Im}\left[ E-E_\text{exc}(\B p)-\Sigma(E, \B p)+i \gamma/2 \right]^{-1}$ as a function of energy $E$ and momentum $\B p$, having introduced a momentum-independent broadening $\gamma$. The polaron states appear as peaks in the exciton spectral function, with the lower and higher energy peak corresponding to the attractive and repulsive polaron, respectively.  The attractive polaron is bound to a dressing cloud of electrons, which lowers the energy below that of the bare exciton (dashed line). It also inherits the roton minimum from the exciton as shown in Fig.~\ref{fig:fig2}(b), (c). While the range of validity of the approximations employed here might not extend to the highest electron densities considered, we expect the roton minimum to be a robust feature of the polaron dispersion.

  The reflection, transmission, and absorption by the monolayer is determined in linear response by the optical conductivity~\cite{efimkin2017many-body}
  \begin{equation}
    \sigma(E) \propto \left| \sum_{\B k} \varphi_0(\B k) \right|^2 S(E, 0).
    \label{eq:conductivity}
  \end{equation}
  The in-plane momentum is zero since the momentum transfered by a photon is negligible for all incident angles. The optical conductivity is therefore insensitive to the presence or absence of the roton state. This is illustrated in \figref{fig:fig2}(d), where we plot the optical conductivity as a function of Fermi energy. At high doping, the spectral weight is transferred from the repulsive polaron to the attractive polaron at zero momentum rather than the lowest energy state corresponding to the roton minimum (dashed line).

\begin{figure}
    \includegraphics[width=\columnwidth]{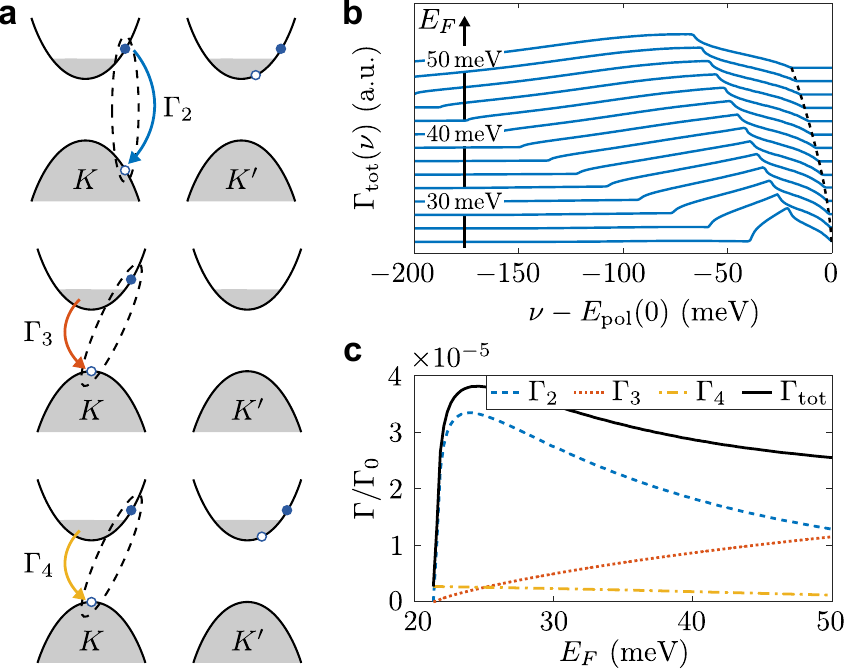}
    \caption{(a) Illustration of different polaron decay channels. (b) Emission spectra computed for the polaronic roton state at different Fermi energies. The dashed line indicates the energy of the roton state, where zero energy is defined by the attractive polaron at zero momentum. (c) Radiative recombination rates $\Gamma_i$ of the polaronic roton state as a function of Fermi energy. Here, $\Gamma_0$ is the decay rate of an exciton at zero momentum and zero Fermi energy.}
    \label{fig:fig3}
  \end{figure}

  In contrast to the optical conductivity, the roton state drastically modifies the photoluminescence properties of MoSe$_2$. To compute the emission spectrum and radiative decay rates, we apply Fermi's Golden Rule to the fully quantized light-matter interaction Hamiltonian using the polaronic wavefunction obtained from the Chevy ansatz as the initial state. This yields the spectral emission rate $\Gamma_\text{tot}(\nu)$ of a photon with frequency $\nu$, which may be split into four separate terms, $\Gamma_\text{tot}(\nu) = \sum_{i = 1}^4 \Gamma_i(\nu)$, according to four different physical processes (see~\cite{SupMat}). The first rate, $\Gamma_1$, corresponds to decay of the pure excitonic component of the polaron (amplitude $\alpha_{\B p}$ in $b_{\B p}^\dagger$). For a zero momentum state, it is sharply peaked at $\nu = E_\text{pol}(0)$ with a magnitude comparable to the decay rate $\Gamma_0$ of an exciton at zero momentum in undoped MoSe$_2$. The rate vanishes if $|\B p| > \nu / c$ and the decay is instead determined by the remaining three rates, whose underlying process is represented schematically in \figref{fig:fig3}(a). The process associated with $\Gamma_2$ is similar to $\Gamma_1$ as it involves the recombination of the correlated electron-hole pair forming the exciton but it does so while leaving behind an electron-hole pair in the opposite Fermi sea. The rates $\Gamma_3$ and $\Gamma_4$ result from radiative recombination of the valence band hole with an electron from the Fermi sea.

  In \figref{fig:fig3}(b), we plot the emission spectrum $\Gamma_\text{tot}(\nu)$ of the roton-like polaronic ground state at different Fermi energies. The zero of the frequency axis is chosen to coincide with the energy of the attractive polaron at zero momentum, where maximum absorption is expected. The emission is clearly redshifted relative to absorption. In fact, the emission peak occurs an energy of roughly $E_F$ below the roton state (dashed line) as a result of the electron-hole pair excitations left behind in the Fermi sea. The spectrally integrated decay rates, defined by $\Gamma_i = \int \di \nu \, \Gamma_i(\nu)$, are shown in \figref{fig:fig3}(c). 

  Phonons may significantly contribute to the decay rate since they offer an additional pathway for the roton-like state to deposit its excess momentum. However, we expect that phonon-assisted decay gives rise to a qualitatively similar redshift and broadening as the decay channels discussed above. A complete calculation of the photoluminescence (PL) spectrum must further take into account the distribution of occupied states, which depends on pumping and relaxation mechanisms~\cite{robert2016exciton}. A detailed understanding of the relevant processes in TMDs is currently missing. 
  
  Our results offer a potential explanation for experimental observations in absorption and PL spectra~\cite{wang2017probing}. In particular, at high electron densities, the emission spectrum begins to redshift relative to the absorption peak and concurrently broadens and decreases in intensity. Both the energy shift and the broadening scale linearly with the Fermi energy $E_F$, with a proportionality constant of order unity. This doping dependent splitting between absorption and PL with an onset at a finite electron density is consistent with our model, where absorption probes zero momentum states, while PL can originate from states close to the lowest energy polaron. Additional experiments may provide further verification of the proposed mechanism. A measurement of the sign change of the polaron mass at zero momentum would provide direct evidence for the emergence of a roton state. Due to the small photon momentum, such measurements are challenging although near-field techniques~\cite{dunn1999near} or acoustic waves~\cite{schiefele2013coupling} could allow one to couple to polaron states with momenta exceeding $\nu / c$. PL lifetime measurements offer a simpler but more indirect probe of the polaron dispersion. Our above treatment suggests that the PL lifetime is drastically increased when the roton minimum is fully developed. 
  
In summary, we proposed a mechanism for the formation of roton states of optical exciations in TMD monolayers. Besides playing an important role in determining optical properties, the roton states open up exciting avenues for creating exotic phases of optical excitations such as supersolids~\cite{henkel2010three-dimensional} or chiral spin liquids~\cite{sedrakyan2015spontaneous}. In particular, if the optical excitations condense into the roton-like minimum, the exact form of interactions between excitations determines whether it is energetically favorable for the condensate to occupy a single momentum state or to fragment~\cite{henkel2010three-dimensional}. The latter scenario gives rise to a superfluid state with periodic density modulation, corresponding to a supersolid. Future work may extend the quantititative validity of our approach beyond low electron densities by including backflow corrections, adding a treatment of dynamical screening, and relaxing the rigid exciton approximation.

\begin{acknowledgments}
  We thank R.~Schmidt and E.~A.~Demler for insightful discussions. This work was supported by an ERC Advanced Investigator Grant (POLTDES), a grant from SNF, the NSF Center for Ultracold Atoms, AFOSR, and the V. Bush Faculty Fellowship. OC and DSW contributed equally to this work. 
\end{acknowledgments}

\bibliography{references}

\begin{thebibliography}{40}%
\makeatletter
\providecommand \@ifxundefined [1]{%
 \@ifx{#1\undefined}
}%
\providecommand \@ifnum [1]{%
 \ifnum #1\expandafter \@firstoftwo
 \else \expandafter \@secondoftwo
 \fi
}%
\providecommand \@ifx [1]{%
 \ifx #1\expandafter \@firstoftwo
 \else \expandafter \@secondoftwo
 \fi
}%
\providecommand \natexlab [1]{#1}%
\providecommand \enquote  [1]{``#1''}%
\providecommand \bibnamefont  [1]{#1}%
\providecommand \bibfnamefont [1]{#1}%
\providecommand \citenamefont [1]{#1}%
\providecommand \href@noop [0]{\@secondoftwo}%
\providecommand \href [0]{\begingroup \@sanitize@url \@href}%
\providecommand \@href[1]{\@@startlink{#1}\@@href}%
\providecommand \@@href[1]{\endgroup#1\@@endlink}%
\providecommand \@sanitize@url [0]{\catcode `\\12\catcode `\$12\catcode
  `\&12\catcode `\#12\catcode `\^12\catcode `\_12\catcode `\%12\relax}%
\providecommand \@@startlink[1]{}%
\providecommand \@@endlink[0]{}%
\providecommand \url  [0]{\begingroup\@sanitize@url \@url }%
\providecommand \@url [1]{\endgroup\@href {#1}{\urlprefix }}%
\providecommand \urlprefix  [0]{URL }%
\providecommand \Eprint [0]{\href }%
\providecommand \doibase [0]{http://dx.doi.org/}%
\providecommand \selectlanguage [0]{\@gobble}%
\providecommand \bibinfo  [0]{\@secondoftwo}%
\providecommand \bibfield  [0]{\@secondoftwo}%
\providecommand \translation [1]{[#1]}%
\providecommand \BibitemOpen [0]{}%
\providecommand \bibitemStop [0]{}%
\providecommand \bibitemNoStop [0]{.\EOS\space}%
\providecommand \EOS [0]{\spacefactor3000\relax}%
\providecommand \BibitemShut  [1]{\csname bibitem#1\endcsname}%
\let\auto@bib@innerbib\@empty
\bibitem [{\citenamefont {Landau}(1941)}]{landau1941theory}%
  \BibitemOpen
  \bibfield  {author} {\bibinfo {author} {\bibfnamefont {L.}~\bibnamefont
  {Landau}},\ }\href {\doibase 10.1103/PhysRev.60.356} {\bibfield  {journal}
  {\bibinfo  {journal} {Phys. Rev.}\ }\textbf {\bibinfo {volume} {60}},\
  \bibinfo {pages} {356} (\bibinfo {year} {1941})}\BibitemShut {NoStop}%
\bibitem [{\citenamefont {Henshaw}\ and\ \citenamefont
  {Woods}(1961)}]{henshaw1961modes}%
  \BibitemOpen
  \bibfield  {author} {\bibinfo {author} {\bibfnamefont {D.~G.}\ \bibnamefont
  {Henshaw}}\ and\ \bibinfo {author} {\bibfnamefont {A.~D.~B.}\ \bibnamefont
  {Woods}},\ }\href {\doibase 10.1103/PhysRev.121.1266} {\bibfield  {journal}
  {\bibinfo  {journal} {Phys. Rev.}\ }\textbf {\bibinfo {volume} {121}},\
  \bibinfo {pages} {1266} (\bibinfo {year} {1961})}\BibitemShut {NoStop}%
\bibitem [{\citenamefont {Mottl}\ \emph {et~al.}(2012)\citenamefont {Mottl},
  \citenamefont {Brennecke}, \citenamefont {Baumann}, \citenamefont {Landig},
  \citenamefont {Donner},\ and\ \citenamefont {Esslinger}}]{mottl2012roton}%
  \BibitemOpen
  \bibfield  {author} {\bibinfo {author} {\bibfnamefont {R.}~\bibnamefont
  {Mottl}}, \bibinfo {author} {\bibfnamefont {F.}~\bibnamefont {Brennecke}},
  \bibinfo {author} {\bibfnamefont {K.}~\bibnamefont {Baumann}}, \bibinfo
  {author} {\bibfnamefont {R.}~\bibnamefont {Landig}}, \bibinfo {author}
  {\bibfnamefont {T.}~\bibnamefont {Donner}}, \ and\ \bibinfo {author}
  {\bibfnamefont {T.}~\bibnamefont {Esslinger}},\ }\href {\doibase
  10.1126/science.1220314} {\bibfield  {journal} {\bibinfo  {journal}
  {Science}\ }\textbf {\bibinfo {volume} {336}},\ \bibinfo {pages} {1570}
  (\bibinfo {year} {2012})}\BibitemShut {NoStop}%
\bibitem [{\citenamefont {Chomaz}\ \emph {et~al.}(2018)\citenamefont {Chomaz},
  \citenamefont {van Bijnen}, \citenamefont {Petter}, \citenamefont {Faraoni},
  \citenamefont {Baier}, \citenamefont {Becher}, \citenamefont {Mark},
  \citenamefont {W{\"{a}}chtler}, \citenamefont {Santos},\ and\ \citenamefont
  {Ferlaino}}]{chomaz2018observation}%
  \BibitemOpen
  \bibfield  {author} {\bibinfo {author} {\bibfnamefont {L.}~\bibnamefont
  {Chomaz}}, \bibinfo {author} {\bibfnamefont {R.~M.~W.}\ \bibnamefont {van
  Bijnen}}, \bibinfo {author} {\bibfnamefont {D.}~\bibnamefont {Petter}},
  \bibinfo {author} {\bibfnamefont {G.}~\bibnamefont {Faraoni}}, \bibinfo
  {author} {\bibfnamefont {S.}~\bibnamefont {Baier}}, \bibinfo {author}
  {\bibfnamefont {J.~H.}\ \bibnamefont {Becher}}, \bibinfo {author}
  {\bibfnamefont {M.~J.}\ \bibnamefont {Mark}}, \bibinfo {author}
  {\bibfnamefont {F.}~\bibnamefont {W{\"{a}}chtler}}, \bibinfo {author}
  {\bibfnamefont {L.}~\bibnamefont {Santos}}, \ and\ \bibinfo {author}
  {\bibfnamefont {F.}~\bibnamefont {Ferlaino}},\ }\href {\doibase
  10.1038/s41567-018-0054-7} {\bibfield  {journal} {\bibinfo  {journal} {Nat.
  Phys.}\ }\textbf {\bibinfo {volume} {14}},\ \bibinfo {pages} {442} (\bibinfo
  {year} {2018})}\BibitemShut {NoStop}%
\bibitem [{\citenamefont {Khamehchi}\ \emph {et~al.}(2014)\citenamefont
  {Khamehchi}, \citenamefont {Zhang}, \citenamefont {Hamner}, \citenamefont
  {Busch},\ and\ \citenamefont {Engels}}]{khamehchi2014measurement}%
  \BibitemOpen
  \bibfield  {author} {\bibinfo {author} {\bibfnamefont {M.~A.}\ \bibnamefont
  {Khamehchi}}, \bibinfo {author} {\bibfnamefont {Y.}~\bibnamefont {Zhang}},
  \bibinfo {author} {\bibfnamefont {C.}~\bibnamefont {Hamner}}, \bibinfo
  {author} {\bibfnamefont {T.}~\bibnamefont {Busch}}, \ and\ \bibinfo {author}
  {\bibfnamefont {P.}~\bibnamefont {Engels}},\ }\href {\doibase
  10.1103/PhysRevA.90.063624} {\bibfield  {journal} {\bibinfo  {journal} {Phys.
  Rev. A}\ }\textbf {\bibinfo {volume} {90}},\ \bibinfo {pages} {063624}
  (\bibinfo {year} {2014})}\BibitemShut {NoStop}%
\bibitem [{\citenamefont {Ji}\ \emph {et~al.}(2015)\citenamefont {Ji},
  \citenamefont {Zhang}, \citenamefont {Xu}, \citenamefont {Wu}, \citenamefont
  {Deng}, \citenamefont {Chen},\ and\ \citenamefont {Pan}}]{ji2015softening}%
  \BibitemOpen
  \bibfield  {author} {\bibinfo {author} {\bibfnamefont {S.-C.}\ \bibnamefont
  {Ji}}, \bibinfo {author} {\bibfnamefont {L.}~\bibnamefont {Zhang}}, \bibinfo
  {author} {\bibfnamefont {X.-T.}\ \bibnamefont {Xu}}, \bibinfo {author}
  {\bibfnamefont {Z.}~\bibnamefont {Wu}}, \bibinfo {author} {\bibfnamefont
  {Y.}~\bibnamefont {Deng}}, \bibinfo {author} {\bibfnamefont {S.}~\bibnamefont
  {Chen}}, \ and\ \bibinfo {author} {\bibfnamefont {J.-W.}\ \bibnamefont
  {Pan}},\ }\href {\doibase 10.1103/PhysRevLett.114.105301} {\bibfield
  {journal} {\bibinfo  {journal} {Phys. Rev. Lett.}\ }\textbf {\bibinfo
  {volume} {114}},\ \bibinfo {pages} {105301} (\bibinfo {year}
  {2015})}\BibitemShut {NoStop}%
\bibitem [{\citenamefont {Ha}\ \emph {et~al.}(2015)\citenamefont {Ha},
  \citenamefont {Clark}, \citenamefont {Parker}, \citenamefont {Anderson},\
  and\ \citenamefont {Chin}}]{ha2015roton-maxon}%
  \BibitemOpen
  \bibfield  {author} {\bibinfo {author} {\bibfnamefont {L.-C.}\ \bibnamefont
  {Ha}}, \bibinfo {author} {\bibfnamefont {L.~W.}\ \bibnamefont {Clark}},
  \bibinfo {author} {\bibfnamefont {C.~V.}\ \bibnamefont {Parker}}, \bibinfo
  {author} {\bibfnamefont {B.~M.}\ \bibnamefont {Anderson}}, \ and\ \bibinfo
  {author} {\bibfnamefont {C.}~\bibnamefont {Chin}},\ }\href {\doibase
  10.1103/PhysRevLett.114.055301} {\bibfield  {journal} {\bibinfo  {journal}
  {Phys. Rev. Lett.}\ }\textbf {\bibinfo {volume} {114}},\ \bibinfo {pages}
  {055301} (\bibinfo {year} {2015})}\BibitemShut {NoStop}%
\bibitem [{\citenamefont {Mak}\ and\ \citenamefont
  {Shan}(2016)}]{mak2016photonics}%
  \BibitemOpen
  \bibfield  {author} {\bibinfo {author} {\bibfnamefont {K.~F.}\ \bibnamefont
  {Mak}}\ and\ \bibinfo {author} {\bibfnamefont {J.}~\bibnamefont {Shan}},\
  }\href {\doibase 10.1038/nphoton.2015.282} {\bibfield  {journal} {\bibinfo
  {journal} {Nat. Photonics}\ }\textbf {\bibinfo {volume} {10}},\ \bibinfo
  {pages} {216} (\bibinfo {year} {2016})}\BibitemShut {NoStop}%
\bibitem [{\citenamefont {Geim}\ and\ \citenamefont
  {Grigorieva}(2013)}]{geim2013van}%
  \BibitemOpen
  \bibfield  {author} {\bibinfo {author} {\bibfnamefont {A.~K.}\ \bibnamefont
  {Geim}}\ and\ \bibinfo {author} {\bibfnamefont {I.~V.}\ \bibnamefont
  {Grigorieva}},\ }\href {\doibase 10.1038/nature12385} {\bibfield  {journal}
  {\bibinfo  {journal} {Nature}\ }\textbf {\bibinfo {volume} {499}},\ \bibinfo
  {pages} {419} (\bibinfo {year} {2013})}\BibitemShut {NoStop}%
\bibitem [{\citenamefont {Raja}\ \emph {et~al.}(2017)\citenamefont {Raja},
  \citenamefont {Chaves}, \citenamefont {Yu}, \citenamefont {Arefe},
  \citenamefont {Hill}, \citenamefont {Rigosi}, \citenamefont {Berkelbach},
  \citenamefont {Nagler}, \citenamefont {Sch{\"{u}}ller}, \citenamefont {Korn},
  \citenamefont {Nuckolls}, \citenamefont {Hone}, \citenamefont {Brus},
  \citenamefont {Heinz}, \citenamefont {Reichman},\ and\ \citenamefont
  {Chernikov}}]{raja2017coulomb}%
  \BibitemOpen
  \bibfield  {author} {\bibinfo {author} {\bibfnamefont {A.}~\bibnamefont
  {Raja}}, \bibinfo {author} {\bibfnamefont {A.}~\bibnamefont {Chaves}},
  \bibinfo {author} {\bibfnamefont {J.}~\bibnamefont {Yu}}, \bibinfo {author}
  {\bibfnamefont {G.}~\bibnamefont {Arefe}}, \bibinfo {author} {\bibfnamefont
  {H.~M.}\ \bibnamefont {Hill}}, \bibinfo {author} {\bibfnamefont {A.~F.}\
  \bibnamefont {Rigosi}}, \bibinfo {author} {\bibfnamefont {T.~C.}\
  \bibnamefont {Berkelbach}}, \bibinfo {author} {\bibfnamefont
  {P.}~\bibnamefont {Nagler}}, \bibinfo {author} {\bibfnamefont
  {C.}~\bibnamefont {Sch{\"{u}}ller}}, \bibinfo {author} {\bibfnamefont
  {T.}~\bibnamefont {Korn}}, \bibinfo {author} {\bibfnamefont {C.}~\bibnamefont
  {Nuckolls}}, \bibinfo {author} {\bibfnamefont {J.}~\bibnamefont {Hone}},
  \bibinfo {author} {\bibfnamefont {L.~E.}\ \bibnamefont {Brus}}, \bibinfo
  {author} {\bibfnamefont {T.~F.}\ \bibnamefont {Heinz}}, \bibinfo {author}
  {\bibfnamefont {D.~R.}\ \bibnamefont {Reichman}}, \ and\ \bibinfo {author}
  {\bibfnamefont {A.}~\bibnamefont {Chernikov}},\ }\href {\doibase
  10.1038/ncomms15251} {\bibfield  {journal} {\bibinfo  {journal} {Nat.
  Commun.}\ }\textbf {\bibinfo {volume} {8}},\ \bibinfo {pages} {15251}
  (\bibinfo {year} {2017})}\BibitemShut {NoStop}%
\bibitem [{\citenamefont {Back}\ \emph {et~al.}(2017)\citenamefont {Back},
  \citenamefont {Sidler}, \citenamefont {Cotlet}, \citenamefont {Srivastava},
  \citenamefont {Takemura}, \citenamefont {Kroner},\ and\ \citenamefont
  {Imamo{\u{g}}lu}}]{back2017giant}%
  \BibitemOpen
  \bibfield  {author} {\bibinfo {author} {\bibfnamefont {P.}~\bibnamefont
  {Back}}, \bibinfo {author} {\bibfnamefont {M.}~\bibnamefont {Sidler}},
  \bibinfo {author} {\bibfnamefont {O.}~\bibnamefont {Cotlet}}, \bibinfo
  {author} {\bibfnamefont {A.}~\bibnamefont {Srivastava}}, \bibinfo {author}
  {\bibfnamefont {N.}~\bibnamefont {Takemura}}, \bibinfo {author}
  {\bibfnamefont {M.}~\bibnamefont {Kroner}}, \ and\ \bibinfo {author}
  {\bibfnamefont {A.}~\bibnamefont {Imamo{\u{g}}lu}},\ }\href {\doibase
  10.1103/PhysRevLett.118.237404} {\bibfield  {journal} {\bibinfo  {journal}
  {Phys. Rev. Lett.}\ }\textbf {\bibinfo {volume} {118}},\ \bibinfo {pages}
  {237404} (\bibinfo {year} {2017})}\BibitemShut {NoStop}%
\bibitem [{\citenamefont {Wang}\ \emph {et~al.}(2018)\citenamefont {Wang},
  \citenamefont {Mak},\ and\ \citenamefont {Shan}}]{wang2018strongly}%
  \BibitemOpen
  \bibfield  {author} {\bibinfo {author} {\bibfnamefont {Z.}~\bibnamefont
  {Wang}}, \bibinfo {author} {\bibfnamefont {K.~F.}\ \bibnamefont {Mak}}, \
  and\ \bibinfo {author} {\bibfnamefont {J.}~\bibnamefont {Shan}},\ }\href
  {\doibase 10.1103/PhysRevLett.120.066402} {\bibfield  {journal} {\bibinfo
  {journal} {Phys. Rev. Lett.}\ }\textbf {\bibinfo {volume} {120}},\ \bibinfo
  {pages} {066402} (\bibinfo {year} {2018})}\BibitemShut {NoStop}%
\bibitem [{\citenamefont {Massignan}\ \emph {et~al.}(2014)\citenamefont
  {Massignan}, \citenamefont {Zaccanti},\ and\ \citenamefont
  {Bruun}}]{massignan2014polarons}%
  \BibitemOpen
  \bibfield  {author} {\bibinfo {author} {\bibfnamefont {P.}~\bibnamefont
  {Massignan}}, \bibinfo {author} {\bibfnamefont {M.}~\bibnamefont {Zaccanti}},
  \ and\ \bibinfo {author} {\bibfnamefont {G.~M.}\ \bibnamefont {Bruun}},\
  }\href {\doibase 10.1088/0034-4885/77/3/034401} {\bibfield  {journal}
  {\bibinfo  {journal} {Reports Prog. Phys.}\ }\textbf {\bibinfo {volume}
  {77}},\ \bibinfo {pages} {034401} (\bibinfo {year} {2014})}\BibitemShut
  {NoStop}%
\bibitem [{\citenamefont {Parish}\ and\ \citenamefont
  {Levinsen}(2013)}]{parish2013highly}%
  \BibitemOpen
  \bibfield  {author} {\bibinfo {author} {\bibfnamefont {M.~M.}\ \bibnamefont
  {Parish}}\ and\ \bibinfo {author} {\bibfnamefont {J.}~\bibnamefont
  {Levinsen}},\ }\href {\doibase 10.1103/PhysRevA.87.033616} {\bibfield
  {journal} {\bibinfo  {journal} {Phys. Rev. A}\ }\textbf {\bibinfo {volume}
  {87}},\ \bibinfo {pages} {033616} (\bibinfo {year} {2013})}\BibitemShut
  {NoStop}%
\bibitem [{\citenamefont {Sidler}\ \emph {et~al.}(2016)\citenamefont {Sidler},
  \citenamefont {Back}, \citenamefont {Cotlet}, \citenamefont {Srivastava},
  \citenamefont {Fink}, \citenamefont {Kroner}, \citenamefont {Demler},\ and\
  \citenamefont {Imamoglu}}]{sidler2016fermi}%
  \BibitemOpen
  \bibfield  {author} {\bibinfo {author} {\bibfnamefont {M.}~\bibnamefont
  {Sidler}}, \bibinfo {author} {\bibfnamefont {P.}~\bibnamefont {Back}},
  \bibinfo {author} {\bibfnamefont {O.}~\bibnamefont {Cotlet}}, \bibinfo
  {author} {\bibfnamefont {A.}~\bibnamefont {Srivastava}}, \bibinfo {author}
  {\bibfnamefont {T.}~\bibnamefont {Fink}}, \bibinfo {author} {\bibfnamefont
  {M.}~\bibnamefont {Kroner}}, \bibinfo {author} {\bibfnamefont
  {E.}~\bibnamefont {Demler}}, \ and\ \bibinfo {author} {\bibfnamefont
  {A.}~\bibnamefont {Imamoglu}},\ }\href {\doibase 10.1038/nphys3949}
  {\bibfield  {journal} {\bibinfo  {journal} {Nat. Phys.}\ }\textbf {\bibinfo
  {volume} {13}},\ \bibinfo {pages} {255} (\bibinfo {year} {2016})}\BibitemShut
  {NoStop}%
\bibitem [{\citenamefont {Efimkin}\ and\ \citenamefont
  {MacDonald}(2017)}]{efimkin2017many-body}%
  \BibitemOpen
  \bibfield  {author} {\bibinfo {author} {\bibfnamefont {D.~K.}\ \bibnamefont
  {Efimkin}}\ and\ \bibinfo {author} {\bibfnamefont {A.~H.}\ \bibnamefont
  {MacDonald}},\ }\href {\doibase 10.1103/PhysRevB.95.035417} {\bibfield
  {journal} {\bibinfo  {journal} {Phys. Rev. B}\ }\textbf {\bibinfo {volume}
  {95}},\ \bibinfo {pages} {035417} (\bibinfo {year} {2017})}\BibitemShut
  {NoStop}%
\bibitem [{\citenamefont {Wang}\ \emph {et~al.}(2017)\citenamefont {Wang},
  \citenamefont {Zhao}, \citenamefont {Mak},\ and\ \citenamefont
  {Shan}}]{wang2017probing}%
  \BibitemOpen
  \bibfield  {author} {\bibinfo {author} {\bibfnamefont {Z.}~\bibnamefont
  {Wang}}, \bibinfo {author} {\bibfnamefont {L.}~\bibnamefont {Zhao}}, \bibinfo
  {author} {\bibfnamefont {K.~F.}\ \bibnamefont {Mak}}, \ and\ \bibinfo
  {author} {\bibfnamefont {J.}~\bibnamefont {Shan}},\ }\href {\doibase
  10.1021/acs.nanolett.6b03855} {\bibfield  {journal} {\bibinfo  {journal}
  {Nano Lett.}\ }\textbf {\bibinfo {volume} {17}},\ \bibinfo {pages} {740}
  (\bibinfo {year} {2017})}\BibitemShut {NoStop}%
\bibitem [{\citenamefont {Yu}\ \emph {et~al.}(2014)\citenamefont {Yu},
  \citenamefont {Liu}, \citenamefont {Gong}, \citenamefont {Xu},\ and\
  \citenamefont {Yao}}]{yu2014dirac}%
  \BibitemOpen
  \bibfield  {author} {\bibinfo {author} {\bibfnamefont {H.}~\bibnamefont
  {Yu}}, \bibinfo {author} {\bibfnamefont {G.-B.}\ \bibnamefont {Liu}},
  \bibinfo {author} {\bibfnamefont {P.}~\bibnamefont {Gong}}, \bibinfo {author}
  {\bibfnamefont {X.}~\bibnamefont {Xu}}, \ and\ \bibinfo {author}
  {\bibfnamefont {W.}~\bibnamefont {Yao}},\ }\href {\doibase
  10.1038/ncomms4876} {\bibfield  {journal} {\bibinfo  {journal} {Nat.
  Commun.}\ }\textbf {\bibinfo {volume} {5}},\ \bibinfo {pages} {3876}
  (\bibinfo {year} {2014})}\BibitemShut {NoStop}%
\bibitem [{\citenamefont {Wu}\ \emph {et~al.}(2015)\citenamefont {Wu},
  \citenamefont {Qu},\ and\ \citenamefont {MacDonald}}]{wu2015exciton}%
  \BibitemOpen
  \bibfield  {author} {\bibinfo {author} {\bibfnamefont {F.}~\bibnamefont
  {Wu}}, \bibinfo {author} {\bibfnamefont {F.}~\bibnamefont {Qu}}, \ and\
  \bibinfo {author} {\bibfnamefont {A.~H.}\ \bibnamefont {MacDonald}},\ }\href
  {\doibase 10.1103/PhysRevB.91.075310} {\bibfield  {journal} {\bibinfo
  {journal} {Phys. Rev. B}\ }\textbf {\bibinfo {volume} {91}},\ \bibinfo
  {pages} {075310} (\bibinfo {year} {2015})}\BibitemShut {NoStop}%
\bibitem [{\citenamefont {Qiu}\ \emph {et~al.}(2015)\citenamefont {Qiu},
  \citenamefont {Cao},\ and\ \citenamefont {Louie}}]{qiu2015nonanalyticity}%
  \BibitemOpen
  \bibfield  {author} {\bibinfo {author} {\bibfnamefont {D.~Y.}\ \bibnamefont
  {Qiu}}, \bibinfo {author} {\bibfnamefont {T.}~\bibnamefont {Cao}}, \ and\
  \bibinfo {author} {\bibfnamefont {S.~G.}\ \bibnamefont {Louie}},\ }\href
  {\doibase 10.1103/PhysRevLett.115.176801} {\bibfield  {journal} {\bibinfo
  {journal} {Phys. Rev. Lett.}\ }\textbf {\bibinfo {volume} {115}},\ \bibinfo
  {pages} {176801} (\bibinfo {year} {2015})}\BibitemShut {NoStop}%
\bibitem [{\citenamefont {Glazov}\ \emph {et~al.}(2014)\citenamefont {Glazov},
  \citenamefont {Amand}, \citenamefont {Marie}, \citenamefont {Lagarde},
  \citenamefont {Bouet},\ and\ \citenamefont {Urbaszek}}]{glazov2014exciton}%
  \BibitemOpen
  \bibfield  {author} {\bibinfo {author} {\bibfnamefont {M.~M.}\ \bibnamefont
  {Glazov}}, \bibinfo {author} {\bibfnamefont {T.}~\bibnamefont {Amand}},
  \bibinfo {author} {\bibfnamefont {X.}~\bibnamefont {Marie}}, \bibinfo
  {author} {\bibfnamefont {D.}~\bibnamefont {Lagarde}}, \bibinfo {author}
  {\bibfnamefont {L.}~\bibnamefont {Bouet}}, \ and\ \bibinfo {author}
  {\bibfnamefont {B.}~\bibnamefont {Urbaszek}},\ }\href {\doibase
  10.1103/PhysRevB.89.201302} {\bibfield  {journal} {\bibinfo  {journal} {Phys.
  Rev. B}\ }\textbf {\bibinfo {volume} {89}},\ \bibinfo {pages} {201302}
  (\bibinfo {year} {2014})}\BibitemShut {NoStop}%
\bibitem [{\citenamefont {Yu}\ and\ \citenamefont {Wu}(2014)}]{yu2014valley}%
  \BibitemOpen
  \bibfield  {author} {\bibinfo {author} {\bibfnamefont {T.}~\bibnamefont
  {Yu}}\ and\ \bibinfo {author} {\bibfnamefont {M.~W.}\ \bibnamefont {Wu}},\
  }\href {\doibase 10.1103/PhysRevB.89.205303} {\bibfield  {journal} {\bibinfo
  {journal} {Phys. Rev. B}\ }\textbf {\bibinfo {volume} {89}},\ \bibinfo
  {pages} {205303} (\bibinfo {year} {2014})}\BibitemShut {NoStop}%
\bibitem [{\citenamefont {Glazov}\ \emph {et~al.}(2017)\citenamefont {Glazov},
  \citenamefont {Golub}, \citenamefont {Wang}, \citenamefont {Marie},
  \citenamefont {Amand},\ and\ \citenamefont {Urbaszek}}]{glazov2017intrinsic}%
  \BibitemOpen
  \bibfield  {author} {\bibinfo {author} {\bibfnamefont {M.~M.}\ \bibnamefont
  {Glazov}}, \bibinfo {author} {\bibfnamefont {L.~E.}\ \bibnamefont {Golub}},
  \bibinfo {author} {\bibfnamefont {G.}~\bibnamefont {Wang}}, \bibinfo {author}
  {\bibfnamefont {X.}~\bibnamefont {Marie}}, \bibinfo {author} {\bibfnamefont
  {T.}~\bibnamefont {Amand}}, \ and\ \bibinfo {author} {\bibfnamefont
  {B.}~\bibnamefont {Urbaszek}},\ }\href {\doibase 10.1103/PhysRevB.95.035311}
  {\bibfield  {journal} {\bibinfo  {journal} {Phys. Rev. B}\ }\textbf {\bibinfo
  {volume} {95}},\ \bibinfo {pages} {035311} (\bibinfo {year}
  {2017})}\BibitemShut {NoStop}%
\bibitem [{\citenamefont {Berkelbach}\ \emph {et~al.}(2013)\citenamefont
  {Berkelbach}, \citenamefont {Hybertsen},\ and\ \citenamefont
  {Reichman}}]{berkelbach2013theory}%
  \BibitemOpen
  \bibfield  {author} {\bibinfo {author} {\bibfnamefont {T.~C.}\ \bibnamefont
  {Berkelbach}}, \bibinfo {author} {\bibfnamefont {M.~S.}\ \bibnamefont
  {Hybertsen}}, \ and\ \bibinfo {author} {\bibfnamefont {D.~R.}\ \bibnamefont
  {Reichman}},\ }\href {\doibase 10.1103/PhysRevB.88.045318} {\bibfield
  {journal} {\bibinfo  {journal} {Phys. Rev. B}\ }\textbf {\bibinfo {volume}
  {88}},\ \bibinfo {pages} {045318} (\bibinfo {year} {2013})}\BibitemShut
  {NoStop}%
\bibitem [{Sup()}]{SupMat}%
  \BibitemOpen
  \href@noop {} {}\bibinfo {note} {{See Supplemental Material, which includes
  references~\cite{berkelbach2013theory,kormanyos2015k,haug2009quantum}, for
  computational details and for explicit expressions of the decay rates
  $\Gamma_{1,2,3,4}$.}}\BibitemShut {Stop}%
\bibitem [{\citenamefont {Gao}\ \emph {et~al.}(2016)\citenamefont {Gao},
  \citenamefont {Liang}, \citenamefont {Spataru},\ and\ \citenamefont
  {Yang}}]{gao2016dynamical}%
  \BibitemOpen
  \bibfield  {author} {\bibinfo {author} {\bibfnamefont {S.}~\bibnamefont
  {Gao}}, \bibinfo {author} {\bibfnamefont {Y.}~\bibnamefont {Liang}}, \bibinfo
  {author} {\bibfnamefont {C.~D.}\ \bibnamefont {Spataru}}, \ and\ \bibinfo
  {author} {\bibfnamefont {L.}~\bibnamefont {Yang}},\ }\href {\doibase
  10.1021/acs.nanolett.6b02118} {\bibfield  {journal} {\bibinfo  {journal}
  {Nano Lett.}\ }\textbf {\bibinfo {volume} {16}},\ \bibinfo {pages} {5568}
  (\bibinfo {year} {2016})}\BibitemShut {NoStop}%
\bibitem [{\citenamefont {{Van Tuan}}\ \emph {et~al.}(2017)\citenamefont {{Van
  Tuan}}, \citenamefont {Scharf}, \citenamefont {{\v{Z}}uti{\'{c}}},\ and\
  \citenamefont {Dery}}]{van2017marrying}%
  \BibitemOpen
  \bibfield  {author} {\bibinfo {author} {\bibfnamefont {D.}~\bibnamefont {{Van
  Tuan}}}, \bibinfo {author} {\bibfnamefont {B.}~\bibnamefont {Scharf}},
  \bibinfo {author} {\bibfnamefont {I.}~\bibnamefont {{\v{Z}}uti{\'{c}}}}, \
  and\ \bibinfo {author} {\bibfnamefont {H.}~\bibnamefont {Dery}},\ }\href
  {\doibase 10.1103/PhysRevX.7.041040} {\bibfield  {journal} {\bibinfo
  {journal} {Phys. Rev. X}\ }\textbf {\bibinfo {volume} {7}},\ \bibinfo {pages}
  {041040} (\bibinfo {year} {2017})}\BibitemShut {NoStop}%
\bibitem [{\citenamefont {Yao}\ \emph {et~al.}(2017)\citenamefont {Yao},
  \citenamefont {Yan}, \citenamefont {Kahn}, \citenamefont {Suslu},
  \citenamefont {Liang}, \citenamefont {Barnard}, \citenamefont {Tongay},
  \citenamefont {Zettl}, \citenamefont {Borys},\ and\ \citenamefont
  {Schuck}}]{yao2017optically}%
  \BibitemOpen
  \bibfield  {author} {\bibinfo {author} {\bibfnamefont {K.}~\bibnamefont
  {Yao}}, \bibinfo {author} {\bibfnamefont {A.}~\bibnamefont {Yan}}, \bibinfo
  {author} {\bibfnamefont {S.}~\bibnamefont {Kahn}}, \bibinfo {author}
  {\bibfnamefont {A.}~\bibnamefont {Suslu}}, \bibinfo {author} {\bibfnamefont
  {Y.}~\bibnamefont {Liang}}, \bibinfo {author} {\bibfnamefont {E.~S.}\
  \bibnamefont {Barnard}}, \bibinfo {author} {\bibfnamefont {S.}~\bibnamefont
  {Tongay}}, \bibinfo {author} {\bibfnamefont {A.}~\bibnamefont {Zettl}},
  \bibinfo {author} {\bibfnamefont {N.~J.}\ \bibnamefont {Borys}}, \ and\
  \bibinfo {author} {\bibfnamefont {P.~J.}\ \bibnamefont {Schuck}},\ }\href
  {\doibase 10.1103/PhysRevLett.119.087401} {\bibfield  {journal} {\bibinfo
  {journal} {Phys. Rev. Lett.}\ }\textbf {\bibinfo {volume} {119}},\ \bibinfo
  {pages} {087401} (\bibinfo {year} {2017})}\BibitemShut {NoStop}%
\bibitem [{\citenamefont {Parish}(2011)}]{parish2011polaron-molecule}%
  \BibitemOpen
  \bibfield  {author} {\bibinfo {author} {\bibfnamefont {M.~M.}\ \bibnamefont
  {Parish}},\ }\href {\doibase 10.1103/PhysRevA.83.051603} {\bibfield
  {journal} {\bibinfo  {journal} {Phys. Rev. A}\ }\textbf {\bibinfo {volume}
  {83}},\ \bibinfo {pages} {051603} (\bibinfo {year} {2011})}\BibitemShut
  {NoStop}%
\bibitem [{\citenamefont {Feynman}\ and\ \citenamefont
  {Cohen}(1956)}]{feynman1956energy}%
  \BibitemOpen
  \bibfield  {author} {\bibinfo {author} {\bibfnamefont {R.~P.}\ \bibnamefont
  {Feynman}}\ and\ \bibinfo {author} {\bibfnamefont {M.}~\bibnamefont
  {Cohen}},\ }\href {\doibase 10.1103/PhysRev.102.1189} {\bibfield  {journal}
  {\bibinfo  {journal} {Phys. Rev.}\ }\textbf {\bibinfo {volume} {102}},\
  \bibinfo {pages} {1189} (\bibinfo {year} {1956})}\BibitemShut {NoStop}%
\bibitem [{\citenamefont {Ross}\ \emph {et~al.}(2013)\citenamefont {Ross},
  \citenamefont {Wu}, \citenamefont {Yu}, \citenamefont {Ghimire},
  \citenamefont {Jones}, \citenamefont {Aivazian}, \citenamefont {Yan},
  \citenamefont {Mandrus}, \citenamefont {Xiao}, \citenamefont {Yao},\ and\
  \citenamefont {Xu}}]{ross2013electrical}%
  \BibitemOpen
  \bibfield  {author} {\bibinfo {author} {\bibfnamefont {J.~S.}\ \bibnamefont
  {Ross}}, \bibinfo {author} {\bibfnamefont {S.}~\bibnamefont {Wu}}, \bibinfo
  {author} {\bibfnamefont {H.}~\bibnamefont {Yu}}, \bibinfo {author}
  {\bibfnamefont {N.~J.}\ \bibnamefont {Ghimire}}, \bibinfo {author}
  {\bibfnamefont {A.~M.}\ \bibnamefont {Jones}}, \bibinfo {author}
  {\bibfnamefont {G.}~\bibnamefont {Aivazian}}, \bibinfo {author}
  {\bibfnamefont {J.}~\bibnamefont {Yan}}, \bibinfo {author} {\bibfnamefont
  {D.~G.}\ \bibnamefont {Mandrus}}, \bibinfo {author} {\bibfnamefont
  {D.}~\bibnamefont {Xiao}}, \bibinfo {author} {\bibfnamefont {W.}~\bibnamefont
  {Yao}}, \ and\ \bibinfo {author} {\bibfnamefont {X.}~\bibnamefont {Xu}},\
  }\href {\doibase 10.1038/ncomms2498} {\bibfield  {journal} {\bibinfo
  {journal} {Nat. Commun.}\ }\textbf {\bibinfo {volume} {4}},\ \bibinfo {pages}
  {1474} (\bibinfo {year} {2013})}\BibitemShut {NoStop}%
\bibitem [{\citenamefont {Schmidt}\ \emph {et~al.}(2012)\citenamefont
  {Schmidt}, \citenamefont {Enss}, \citenamefont {Pietil{\"{a}}},\ and\
  \citenamefont {Demler}}]{schmidt2012fermi}%
  \BibitemOpen
  \bibfield  {author} {\bibinfo {author} {\bibfnamefont {R.}~\bibnamefont
  {Schmidt}}, \bibinfo {author} {\bibfnamefont {T.}~\bibnamefont {Enss}},
  \bibinfo {author} {\bibfnamefont {V.}~\bibnamefont {Pietil{\"{a}}}}, \ and\
  \bibinfo {author} {\bibfnamefont {E.}~\bibnamefont {Demler}},\ }\href
  {\doibase 10.1103/PhysRevA.85.021602} {\bibfield  {journal} {\bibinfo
  {journal} {Phys. Rev. A}\ }\textbf {\bibinfo {volume} {85}},\ \bibinfo
  {pages} {021602} (\bibinfo {year} {2012})}\BibitemShut {NoStop}%
\bibitem [{\citenamefont {Chevy}(2006)}]{chevy2006universal}%
  \BibitemOpen
  \bibfield  {author} {\bibinfo {author} {\bibfnamefont {F.}~\bibnamefont
  {Chevy}},\ }\href {\doibase 10.1103/PhysRevA.74.063628} {\bibfield  {journal}
  {\bibinfo  {journal} {Phys. Rev. A}\ }\textbf {\bibinfo {volume} {74}},\
  \bibinfo {pages} {063628} (\bibinfo {year} {2006})}\BibitemShut {NoStop}%
\bibitem [{\citenamefont {Robert}\ \emph {et~al.}(2016)\citenamefont {Robert},
  \citenamefont {Lagarde}, \citenamefont {Cadiz}, \citenamefont {Wang},
  \citenamefont {Lassagne}, \citenamefont {Amand}, \citenamefont {Balocchi},
  \citenamefont {Renucci}, \citenamefont {Tongay}, \citenamefont {Urbaszek},\
  and\ \citenamefont {Marie}}]{robert2016exciton}%
  \BibitemOpen
  \bibfield  {author} {\bibinfo {author} {\bibfnamefont {C.}~\bibnamefont
  {Robert}}, \bibinfo {author} {\bibfnamefont {D.}~\bibnamefont {Lagarde}},
  \bibinfo {author} {\bibfnamefont {F.}~\bibnamefont {Cadiz}}, \bibinfo
  {author} {\bibfnamefont {G.}~\bibnamefont {Wang}}, \bibinfo {author}
  {\bibfnamefont {B.}~\bibnamefont {Lassagne}}, \bibinfo {author}
  {\bibfnamefont {T.}~\bibnamefont {Amand}}, \bibinfo {author} {\bibfnamefont
  {A.}~\bibnamefont {Balocchi}}, \bibinfo {author} {\bibfnamefont
  {P.}~\bibnamefont {Renucci}}, \bibinfo {author} {\bibfnamefont
  {S.}~\bibnamefont {Tongay}}, \bibinfo {author} {\bibfnamefont
  {B.}~\bibnamefont {Urbaszek}}, \ and\ \bibinfo {author} {\bibfnamefont
  {X.}~\bibnamefont {Marie}},\ }\href {\doibase 10.1103/PhysRevB.93.205423}
  {\bibfield  {journal} {\bibinfo  {journal} {Phys. Rev. B}\ }\textbf {\bibinfo
  {volume} {93}},\ \bibinfo {pages} {205423} (\bibinfo {year}
  {2016})}\BibitemShut {NoStop}%
\bibitem [{\citenamefont {Dunn}(1999)}]{dunn1999near}%
  \BibitemOpen
  \bibfield  {author} {\bibinfo {author} {\bibfnamefont {R.~C.}\ \bibnamefont
  {Dunn}},\ }\href {\doibase 10.1021/cr980130e} {\bibfield  {journal} {\bibinfo
   {journal} {Chem. Rev.}\ }\textbf {\bibinfo {volume} {99}},\ \bibinfo {pages}
  {2891} (\bibinfo {year} {1999})}\BibitemShut {NoStop}%
\bibitem [{\citenamefont {Schiefele}\ \emph {et~al.}(2013)\citenamefont
  {Schiefele}, \citenamefont {Pedr{\'{o}}s}, \citenamefont {Sols},
  \citenamefont {Calle},\ and\ \citenamefont {Guinea}}]{schiefele2013coupling}%
  \BibitemOpen
  \bibfield  {author} {\bibinfo {author} {\bibfnamefont {J.}~\bibnamefont
  {Schiefele}}, \bibinfo {author} {\bibfnamefont {J.}~\bibnamefont
  {Pedr{\'{o}}s}}, \bibinfo {author} {\bibfnamefont {F.}~\bibnamefont {Sols}},
  \bibinfo {author} {\bibfnamefont {F.}~\bibnamefont {Calle}}, \ and\ \bibinfo
  {author} {\bibfnamefont {F.}~\bibnamefont {Guinea}},\ }\href {\doibase
  10.1103/PhysRevLett.111.237405} {\bibfield  {journal} {\bibinfo  {journal}
  {Phys. Rev. Lett.}\ }\textbf {\bibinfo {volume} {111}},\ \bibinfo {pages}
  {237405} (\bibinfo {year} {2013})}\BibitemShut {NoStop}%
\bibitem [{\citenamefont {Henkel}\ \emph {et~al.}(2010)\citenamefont {Henkel},
  \citenamefont {Nath},\ and\ \citenamefont
  {Pohl}}]{henkel2010three-dimensional}%
  \BibitemOpen
  \bibfield  {author} {\bibinfo {author} {\bibfnamefont {N.}~\bibnamefont
  {Henkel}}, \bibinfo {author} {\bibfnamefont {R.}~\bibnamefont {Nath}}, \ and\
  \bibinfo {author} {\bibfnamefont {T.}~\bibnamefont {Pohl}},\ }\href {\doibase
  10.1103/PhysRevLett.104.195302} {\bibfield  {journal} {\bibinfo  {journal}
  {Phys. Rev. Lett.}\ }\textbf {\bibinfo {volume} {104}},\ \bibinfo {pages}
  {195302} (\bibinfo {year} {2010})}\BibitemShut {NoStop}%
\bibitem [{\citenamefont {Sedrakyan}\ \emph {et~al.}(2015)\citenamefont
  {Sedrakyan}, \citenamefont {Glazman},\ and\ \citenamefont
  {Kamenev}}]{sedrakyan2015spontaneous}%
  \BibitemOpen
  \bibfield  {author} {\bibinfo {author} {\bibfnamefont {T.~A.}\ \bibnamefont
  {Sedrakyan}}, \bibinfo {author} {\bibfnamefont {L.~I.}\ \bibnamefont
  {Glazman}}, \ and\ \bibinfo {author} {\bibfnamefont {A.}~\bibnamefont
  {Kamenev}},\ }\href {\doibase 10.1103/PhysRevLett.114.037203} {\bibfield
  {journal} {\bibinfo  {journal} {Phys. Rev. Lett.}\ }\textbf {\bibinfo
  {volume} {114}},\ \bibinfo {pages} {037203} (\bibinfo {year}
  {2015})}\BibitemShut {NoStop}%
\bibitem [{\citenamefont {Korm{\'{a}}nyos}\ \emph {et~al.}(2015)\citenamefont
  {Korm{\'{a}}nyos}, \citenamefont {Burkard}, \citenamefont {Gmitra},
  \citenamefont {Fabian}, \citenamefont {Z{\'{o}}lyomi}, \citenamefont
  {Drummond},\ and\ \citenamefont {Fal'ko}}]{kormanyos2015k}%
  \BibitemOpen
  \bibfield  {author} {\bibinfo {author} {\bibfnamefont {A.}~\bibnamefont
  {Korm{\'{a}}nyos}}, \bibinfo {author} {\bibfnamefont {G.}~\bibnamefont
  {Burkard}}, \bibinfo {author} {\bibfnamefont {M.}~\bibnamefont {Gmitra}},
  \bibinfo {author} {\bibfnamefont {J.}~\bibnamefont {Fabian}}, \bibinfo
  {author} {\bibfnamefont {V.}~\bibnamefont {Z{\'{o}}lyomi}}, \bibinfo {author}
  {\bibfnamefont {N.~D.}\ \bibnamefont {Drummond}}, \ and\ \bibinfo {author}
  {\bibfnamefont {V.}~\bibnamefont {Fal'ko}},\ }\href {\doibase
  10.1088/2053-1583/2/2/022001} {\bibfield  {journal} {\bibinfo  {journal} {2D
  Mater.}\ }\textbf {\bibinfo {volume} {2}},\ \bibinfo {pages} {022001}
  (\bibinfo {year} {2015})}\BibitemShut {NoStop}%
\bibitem [{\citenamefont {Haug}\ and\ \citenamefont
  {Koch}(2009)}]{haug2009quantum}%
  \BibitemOpen
  \bibfield  {author} {\bibinfo {author} {\bibfnamefont {H.}~\bibnamefont
  {Haug}}\ and\ \bibinfo {author} {\bibfnamefont {S.~W.}\ \bibnamefont
  {Koch}},\ }\href@noop {} {\emph {\bibinfo {title} {{Quantum Theory of the
  Optical and Electronic Properties of Semiconductors}}}},\ \bibinfo {edition}
  {5th}\ ed.\ (\bibinfo  {publisher} {World Scientific},\ \bibinfo {address}
  {Singapore},\ \bibinfo {year} {2009})\BibitemShut {NoStop}%
\end{thebibliography}%
\end{document}